\pdfoutput=1
\newcommand*{\ATLASLATEXPATH}{latex/}
\documentclass[UKenglish, cernpreprint, texlive=2016]{\ATLASLATEXPATH atlasdoc}

\usepackage[csquotes=false,default,subfigure]{\ATLASLATEXPATH atlaspackage}
\usepackage{\ATLASLATEXPATH atlasbiblatex}

\usepackage{\ATLASLATEXPATH atlasphysics}

\graphicspath{{logos/}{figures/}}

\usepackage{DiffCombiPaper2017-defs}
\usepackage{bm}

\AtlasTitle{Combined measurement of  differential and total cross sections in the $H \rightarrow \gamma \gamma$ and the $H \rightarrow ZZ^* \rightarrow 4\ell$ decay channels at $\sqrt{s} = 13$~\tev\ with the ATLAS detector}

\AtlasAbstract{%
  A combined measurement of differential and inclusive total cross sections of Higgs boson production is performed using 36.1~fb$^{-1}$ of 13~\tev\ proton--proton collision data produced by the LHC and recorded by the ATLAS detector in 2015 and 2016.
Cross sections are obtained from measured $H \rightarrow \gamma \gamma$ and $H \rightarrow ZZ^* \rightarrow 4\ell$ event yields, which are combined taking into account detector efficiencies, resolution, acceptances and branching fractions.
The total Higgs boson production cross section is measured to be \hbox{57.0$^{+6.0}_{-5.9}$ (stat.) $^{+4.0}_{-3.3}$ (syst.)~pb}, in agreement with the Standard Model prediction. 
Differential cross-section measurements are presented for the Higgs boson transverse momentum distribution, Higgs boson rapidity, number of jets produced together with the Higgs boson, and the transverse momentum of the leading jet. 
The results from the two decay channels are found to be compatible, and their combination agrees with the Standard Model predictions.
}

\author{The ATLAS Collaboration}

\AtlasRefCode{HIGG-2017-11}

\PreprintIdNumber{CERN-EP-2018-080}

\AtlasJournalRef{\PLB 786 (2018) 114}
\AtlasDOI{10.1016/j.physletb.2018.09.019}

\hypersetup{pdftitle={ATLAS document},pdfauthor={The ATLAS Collaboration}}

\begin{document}

\maketitle

\section{Introduction}

Differential cross-section measurements are important studies of Higgs boson production, probing Standard Model (SM) predictions. Deviations from the predictions could be caused by physics beyond the SM~\hbox{\cite{Grazzini:2016paz,HIGG-2015-02}}. Both the ATLAS and CMS collaborations have measured differential cross sections in the \hgg, \hbox{\hfourl} (where $\ell = e,\mu$) and $H \rightarrow WW^* \rightarrow e\nu \mu \nu$ decay channels~\cite{HIGG-2013-10,HIGG-2013-22,HIGG-2014-11,HIGG-2015-04,CMS-HIG-14-016,CMS-HIG-14-028,CMS-HIG-15-010,Sirunyan:2017exp}.

This Letter describes the combination of two fiducial cross-section measurements in the \hgg~\cite{HIGG-2016-21} and \hfourl~\cite{HIGG-2016-25} decay channels, which were obtained using 36.1~\ifb\ of $pp$ collision data produced by the Large Hadron Collider (LHC) in 2015 and 2016 with a centre-of-mass energy of 13~\tev\ and recorded by the ATLAS detector~\cite{PERF-2007-01}. The combined cross section is extracted for the total phase space, increasing the degree of model dependence compared to the individual measurements, which were performed in a fiducial phase space close to the selection criteria for reconstructed events in the detector. Despite the additional systematic uncertainties assigned to the extrapolation to the total phase space, the combination significantly reduces the measurement uncertainty compared to the results in the individual decay channels.

The measured observables include the total production cross section, the Higgs boson's transverse momentum~\ptH, sensitive to perturbative QCD calculations, and the Higgs boson's rapidity \yH, sensitive to the parton distribution functions (PDF). Furthermore the number of jets \njets\ is measured in events with a Higgs boson and jet transverse momentum above 30~\gev, as well as the leading jet's transverse momentum \jetpt. Both the \njets\ and \jetpt\ observables probe the theoretical modelling of high-\pT\ QCD radiation in Higgs boson production. The \njets\ observable is also sensitive to the different Higgs boson production processes~\cite{deFlorian:2016spz}.

The cross sections are obtained from yields measured in the $H \rightarrow \gamma \gamma$ and $H \rightarrow ZZ^* \rightarrow 4\ell$ decay channels, which are combined taking into account detector efficiencies, resolution, acceptances and branching fractions. For each decay channel and each observable, the cross sections can be written as
\begin{align*}
\sigma_i = \frac{N^\mathrm{sig}_i}{L\ \mathcal{B}\ \mathcal{A}_i\ \mathcal{C}_i},
\end{align*}
where $i$ is the iterator over the bins of the observable of interest, $\sigma_i$ is the cross section in bin $i$,  $N^\mathrm{sig}_i$ is the number of measured reconstructed signal events following the analysis selection, $L$ is the integrated luminosity and $\mathcal{B}$ is the branching fraction. The term $\mathcal{C}_i$ is the correction factor from the number of events reconstructed to the number of events at particle level produced in the respective fiducial phase space, and $\mathcal{A}_i$ is the acceptance factor extrapolating from the fiducial to the total phase space contained in the bin of interest. 

Predicted branching ratios and production cross sections are obtained for $m_H=125.09$~\gev~\cite{HIGG-2014-14}, as described in Section~\ref{sec:montecarlo}.
The number of signal events in each bin of a probed observable is extracted in the $H \rightarrow \gamma \gamma$ and $H \rightarrow ZZ^* \rightarrow 4\ell$ channels from fits to the \mgg\ and \mfourl\ invariant mass distributions, respectively.
The signal extraction and the correction factors are discussed in detail in Refs.~\cite{HIGG-2016-21,HIGG-2016-25}. The correction factors are obtained from simulated events, assuming SM Higgs boson production.
In order to harmonize the published \hyy\ fiducial measurement~\cite{HIGG-2016-21} with the \hfourl\ analysis~\cite{HIGG-2016-25},  adjustments were made to the bin boundaries and the uncertainties of the correction factors due to the fractions of different Higgs boson production processes in the \hyy\ decay channel. 
To extrapolate to the total phase space,  acceptance factors and uncertainties are calculated for the combination, as discussed in Section~\ref{sec:acceptance}.
 Section~\ref{sec:stats} presents the combination methodology. The results are discussed in Section~\ref{sec:results}.

\section{Higgs boson Monte Carlo samples, cross sections and branching fractions}
\label{sec:montecarlo}
Predictions of SM Higgs boson production are used in the calculation of the correction and acceptance factors, and are compared to the measured cross sections. 
The Monte Carlo (MC) event generators that were used to simulate gluon--gluon fusion (ggF), vector-boson fusion (VBF), associated Higgs boson production ($VH$, $V=W,Z$), and Higgs boson production in association with a heavy-quark pair (\ttH, \bbH) are listed in Table~\ref{tab:samples}. The accuracy of the calculations and the PDF sets used are also given, with the abbreviations NLO for next-to-leading order, NNLO for next-to-next-to-leading order, and NNLL for next-to-next-to-leading logarithm.
For ggF, VBF, $VH$, \bbH\ in both decay channels and \ttH\ in the \hgg\ decay channel, \textsc{Pythia8}~\cite{pythia,Sjostrand:2007gs} was used for the decay, parton shower, hadronization and multiple parton interactions. For \ttH\ in the \hfourl\ decay channel, \textsc{Herwig++}~\cite{Bahr:2008pv,Bellm:2013hwb} was used.

\begin{table}[h]
\caption{Monte Carlo samples used to simulate Higgs boson production, including the generators, accuracy of calculations in QCD, and PDF sets.}
\label{tab:samples}
\centering
\begin{tabular}{ llll}
  \hline
  \hline
  Process & Generator & Accuracy in QCD & PDF set   \\
  \hline
  ggF & \textsc{Powheg-Box} v2 (NNLOPS)~\cite{powheg1,powheg2,powheg3,Campbell:2012am} & NNLO in \yH~\cite{Hamilton:2013fea}, & PDF4LHC~\cite{Butterworth:2015oua}  \\
  & & \ptH\ consistent with \textsc{HqT}  & \\
  & & (NNLO+NNLL)~\cite{Bozzi:2005wk,deFlorian:2011xf} & \\
  VBF & \textsc{Powheg-Box} v2~\cite{powheg1,powheg2,powheg3,powheg5}  & NLO & PDF4LHC  \\
  \VH & \textsc{Powheg-Box} v2 (\textsc{MiNLO})~\cite{powheg1,powheg2,powheg3,Luisoni:2013kna} & NLO & PDF4LHC \\
  \ttH & \textsc{Madgraph5\_aMC@NLO} (v.2.2.3)~\cite{Alwall:2014hca} & NLO & CT10nlo~\cite{Lai:2010vv} \\
  \bbH & \textsc{Madgraph5\_aMC@NLO} (v.2.3.3)~\cite{Alwall:2014hca,Wiesemann:2014ioa} & NLO & NNPDF23~\cite{Ball:2012cx} \\
\hline
\hline
\end{tabular}
\end{table}

The samples are normalized to the cross-section predictions taken from Refs.~\cite{Dittmaier:2011ti,Dittmaier:2012vm,Heinemeyer:2013tqa,deFlorian:2016spz}. These predictions were obtained assuming a Higgs boson mass of 125.09~\gev~\cite{HIGG-2014-14}  to calculate cross sections and branching ratios. Details are given in Table~\ref{tab:XS}, including the accuracy of the calculations, and the composition of the production modes in the SM. N$^3$LO is the abbreviation for next-to-next-to-next-to-leading order, and EW stands for electroweak.

\begin{table}[h]
\caption{Cross-section predictions used to normalize the MC samples, the accuracy of the calculations (in QCD if not noted otherwise), and the composition of the production modes in the SM.}
\label{tab:XS}
\centering
\begin{tabular}{ lll  }
  \hline
  \hline
  Process & Accuracy & Fraction [\%] \\
  \hline
  \vspace{2pt}
  ggF & N$^3$LO, NLO EW corrections~\cite{Djouadi:1991tka,Dawson:1990zj,Spira:1995rr,Harlander:2002wh,Anastasiou:2002yz,Ravindran:2003um,Anastasiou:2014vaa,Anastasiou:2014lda,Anastasiou:2016cez,Aglietti:2004nj,Actis:2008ug,deFlorian:2012yg,Anastasiou:2012hx,Baglio:2010ae}  & 87.4 \\
  VBF & NLO, NLO EW corrections~\cite{Ciccolini:2007jr,Ciccolini:2007ec,Arnold:2008rz} & ~~6.8\\

  & with approximate NNLO QCD corrections~\cite{Bolzoni:2010xr} & \\
  $VH$ & NNLO~\cite{Han:1991ia,Brein:2003wg}, NLO EW corrections~\cite{Ciccolini:2003jy} & ~~4.1\\
  \ttH & NLO, NLO EW corrections~\cite{Beenakker:2001rj,Beenakker:2002nc,Dawson:2002tg,Dawson:2003zu} & ~~0.9\\
  \bbH\ & five-flavour: NNLO, four-flavour: NLO~\cite{Harlander:2011aa} & ~~0.9\\
\hline
\hline
\end{tabular}
\end{table}

In addition to the NNLOPS sample (see Table~\ref{tab:samples}) scaled to the N$^3$LO cross section with a $K$-factor of 1.1, further SM ggF predictions are compared with the measurements. If not mentioned otherwise, the cross sections predicted by the respective calculations are used. For the comparison with data, the non-ggF Higgs boson production processes are added using the samples and cross sections described above.
\begin{itemize}
\item The \ptH\ distribution is compared with the predictions from \textsc{HRes}~\cite{deFlorian:2012mx, Grazzini:2013mca}, \textsc{RaDISH + NNLOJET}~\cite{Monni:2016ktx}, and \textsc{Madgraph5\_aMC@NLO}. \textsc{HRes} includes resummation to NNLL and  computes fixed-order cross sections for ggF Higgs boson production up to NNLO in QCD. It describes the \ptH\ distribution at NLO. Finite $t$-, $b$-, and $c$-quark masses are included at NLO accuracy. The \textsc{RaDISH + NNLOJET} prediction includes resummation to NNLL and matching to the one-jet NNLO differential spectrum from NNLOJET~\cite{Chen:2014gva,Chen:2016zka}. It includes corrections from the finite $t$- and $b$-quark masses. The predictions from \textsc{Madgraph5\_aMC@NLO} are scaled to the N$^3$LO cross section with a $K$-factor of 1.47. 
  This generator provides NLO accuracy in QCD for zero, one, and two additional jets, merged with the FxFx scheme~\cite{Frederix:2012ps,} and includes the finite top quark mass effects~\cite{Alwall:2014hca,Frederix:2016cnl,Mattelaer:2016gcx}.
\item The \yH\  measurement is compared with predictions from \textsc{Madgraph5\_aMC@NLO} merged with the FxFx scheme and \textsc{SCETlib+MCFM8}~\cite{Ebert:2017uel, SCETlib}, which achieves NNLO+NNLL$'_\varphi$ accuracy\footnote{The prime indicates that important parts of the N$^3$LL (next-to-next-to-next-to-leading logarithm) contribution are included along with the full NNLL corrections and the subscript $\varphi$ indicates that resummation is applied to the gluon form factor.} by applying a resummation of the virtual corrections to the gluon form factor. The underlying NNLO predictions are obtained using MCFM8 with zero-jettiness subtractions~\cite{Boughezal:2016wmq, Gaunt:2015pea}.
\item The \jetpt\ measurement is compared with \textsc{SCETlib}, with  NNLL$'+$NNLO$_0$ accuracy\footnote{NNLO$_0$ refers to the NNLO corrections relative to the LO $gg \rightarrow H$ process with 0 additional partons.}~\cite{Stewart:2013faa,SCETlib}.
\item Multiple predictions exist for different bins of the \njets\ distribution. Considered here are the STWZ-BLPTW prediction~\cite{Stewart:2013faa, Boughezal:2013oha, deFlorian:2016spz}, which includes NNLL$'+$NNLO resummation for the \pt\ of the leading jet, combined with a NLL$'+$NLO resummation for the subleading jet, and the JVE-N$^3$LO prediction~\cite{Banfi:2015pju}, which includes NNLL resummation of the \pt\ of the
leading jet with small-R resummation and is matched to the N$^3$LO total cross section. In addition, predictions from \textsc{Madgraph5\_aMC@NLO}, 
are compared with the full \njets\ distribution. 

\end{itemize}
For ggF, VBF and $VH$, the PDF4LHC set is varied according to its eigenvectors~\cite{Butterworth:2015oua}, and the envelope of the variations is used as the systematic uncertainty. The effect of PDF uncertainties on \ttH\ and \bbH\ is negligible and not included. The renormalization and factorization scales are varied by factors of 2.0 and 0.5. For NNLOPS, instead of the internal scale uncertainties, the same scheme as in Refs.~\cite{HIGG-2016-21,HIGG-2016-25,HIGG-2016-22} is used: four parameters account for uncertainties in the cross sections for events with different jet multiplicities~\cite{deFlorian:2016spz,Stewart:2013faa,Liu:2013hba,Boughezal:2013oha}, and three parameters account for the uncertainties in the modelling of the \ptH\ distributions.

The predicted Higgs boson decay branching ratios are (0.227 $\pm$ 0.007)\%  and (0.0125 $\pm$ 0.0003)\% 
for the \hgg\ and \hfourl\ decays, respectively~\cite{deFlorian:2016spz}. Both branching ratio calculations include the complete NLO QCD and EW corrections. 
For \hfourl, the interference effects between identical final-state fermion pairs are included.
The correlations of the branching ratio uncertainties and the dependence of the predicted branching ratios on the Higgs boson mass are taken into account in the combination. For the \hfourl\ decay channel, which has the larger dependence, this corresponds to a relative variation of $\sim$2\% in the branching ratio when varying the assumed Higgs boson mass by $\pm 0.24$~\gev~\cite{HIGG-2014-14}.

\section{Acceptance correction}
\label{sec:acceptance}
\newcommand{\AtlasCoordFootnote}{%
ATLAS uses a right-handed coordinate system with its origin at the nominal interaction point (IP)
in the centre of the detector and the $z$-axis along the beam pipe.
The $x$-axis points from the IP to the centre of the LHC ring,
and the $y$-axis points upwards.
Cylindrical coordinates $(r,\phi)$ are used in the transverse plane, 
$\phi$ being the azimuthal angle around the $z$-axis.
The pseudorapidity is defined in terms of the polar angle $\theta$ as $\eta = -\ln \tan(\theta/2)$.
Angular distance is measured in units of $\Delta R \equiv \sqrt{(\Delta\eta)^{2} + (\Delta\phi)^{2}}$.}

The acceptance factors that extrapolate at particle-level from the \hgg\ and \hfourl\ fiducial phase space to the full phase space are estimated using the MC samples and cross sections described in Section~\ref{sec:montecarlo}. Their evaluation assumes SM Higgs boson production fractions and a Higgs boson mass of 125~\gev; the 90~\mev\ difference from 125.09~\gev\ has negligible impact on the Higgs boson kinematics and is covered by the systematic uncertainty from the Higgs boson mass measurement. 

In the \hgg\ fiducial phase space~\cite{HIGG-2016-21}, the selected events have two photons with pseudorapidity\footnote{\AtlasCoordFootnote} $|\eta| < 1.37$ or 1.52 $< |\eta| < $ 2.37 and $\pt^{\gamma_1} > 0.35 \mgg$, $\pt^{\gamma_2} > 0.25 \mgg$, where $\pt^{\gamma_{1(2)}}$ refers to the transverse momentum of the (sub)leading photon and $\mgg$ is the invariant mass of the two photons. The photons are required to be isolated: the \pt\ of the system of charged generator-level particles within $\Delta R = 0.2$ of the photon is required to be less than 0.05 times the \pt\ of the photon. 
In the \hfourl\ fiducial phase space~\cite{HIGG-2016-25}, the selected events have four muons, four electrons, or two electrons and two muons.
The three leading leptons are required to have \hbox{\pt$>$ 20, 15, 10~\gev}. The lowest-\pt\ muon (electron) has to fulfil $\pt > 5$ (7)~\gev. The muons have to be within \hbox{$|\eta| < 2.7$} and the electrons within $|\eta| < 2.47$. Following the selection of events in data, requirements are placed on the masses of the two same-flavour opposite-charge pairs, on the $\Delta R$ of any two leptons, and the invariant mass of the four-lepton system, 115~\gev~$<$ \mfourl $<$ 130~\gev.

In the total phase space, the quantities \ptH\ and \yH\ are computed directly from the simulated Higgs boson momentum instead of its decay products, as in the fiducial analyses. 
Simulated particle-level
jets are built from all particles with $c\tau >$ 10 mm excluding neutrinos, electrons and muons that do not originate from hadron decays. Photons are excluded from jet finding if they originate directly from the Higgs boson decay or are radiated off leptons from the Higgs boson decay. 
Jets are reconstructed using the anti-$k_t$ algorithm~\cite{jet-antikt} with a radius parameter $R = 0.4$, and are required to have $\pt > 30$~\gev.

Theory uncertainties in the signal acceptance related to the PDF, higher-order corrections, and the parton shower are considered for the acceptance factors and are correlated between the two channels. Uncertainties due to the PDF and scales are estimated as described in Section~\ref{sec:montecarlo}. Uncertainties due to the parton shower are evaluated by comparing the ggF default showering \progname{Pythia8} with \progname{Herwig7}. The uncertainty is derived from the full difference between the two cases.
The Higgs boson mass is varied within the uncertainty of the ATLAS--CMS combined measurement~\cite{HIGG-2014-14}. To account for model dependence, the fractions of production modes are varied within the uncertainties from the dedicated measurements by the ATLAS and CMS collaborations~\cite{HIGG-2015-07}. For \ttH, the 13~\tev\ ATLAS results are used~\cite{HIGG-2017-02}. The \bbH\ cross section is varied within the uncertainties due to the PDF and higher-order corrections~\cite{deFlorian:2016spz}. The total systematic uncertainties of the acceptance factors range between 0.4\% and 5\%, depending on the observable and bin. The parton shower uncertainty dominates.

The inclusive acceptance factors are 50\% for the \hgg\ channel and 42\% for the \hfourl\ channel (relative to the full phase space of $H \rightarrow ZZ^* \rightarrow 2\ell2\ell'$, where $\ell,\ell' = e$ or $\mu$).  The acceptance is lower for \hfourl\ than for \hgg\ since it is less likely for four leptons to fulfil the fiducial requirements. Figure~\ref{fig:acc} shows the acceptance factors used for the differential observables and their systematic uncertainties.  
The fiducial acceptance falls off steeply as the Higgs boson rapidity increases, as both fiducial definitions include pseudorapidity requirements on the Higgs boson decay products. The fiducial acceptance in the \hyy\ channel as a function of \ptH\ is shaped by the \pt\ selection criteria on the photons.

\begin{figure}[!htbp]
  \centering
  \subfigure[\label{fig:acc_pt}]{\includegraphics[width=0.45\linewidth]{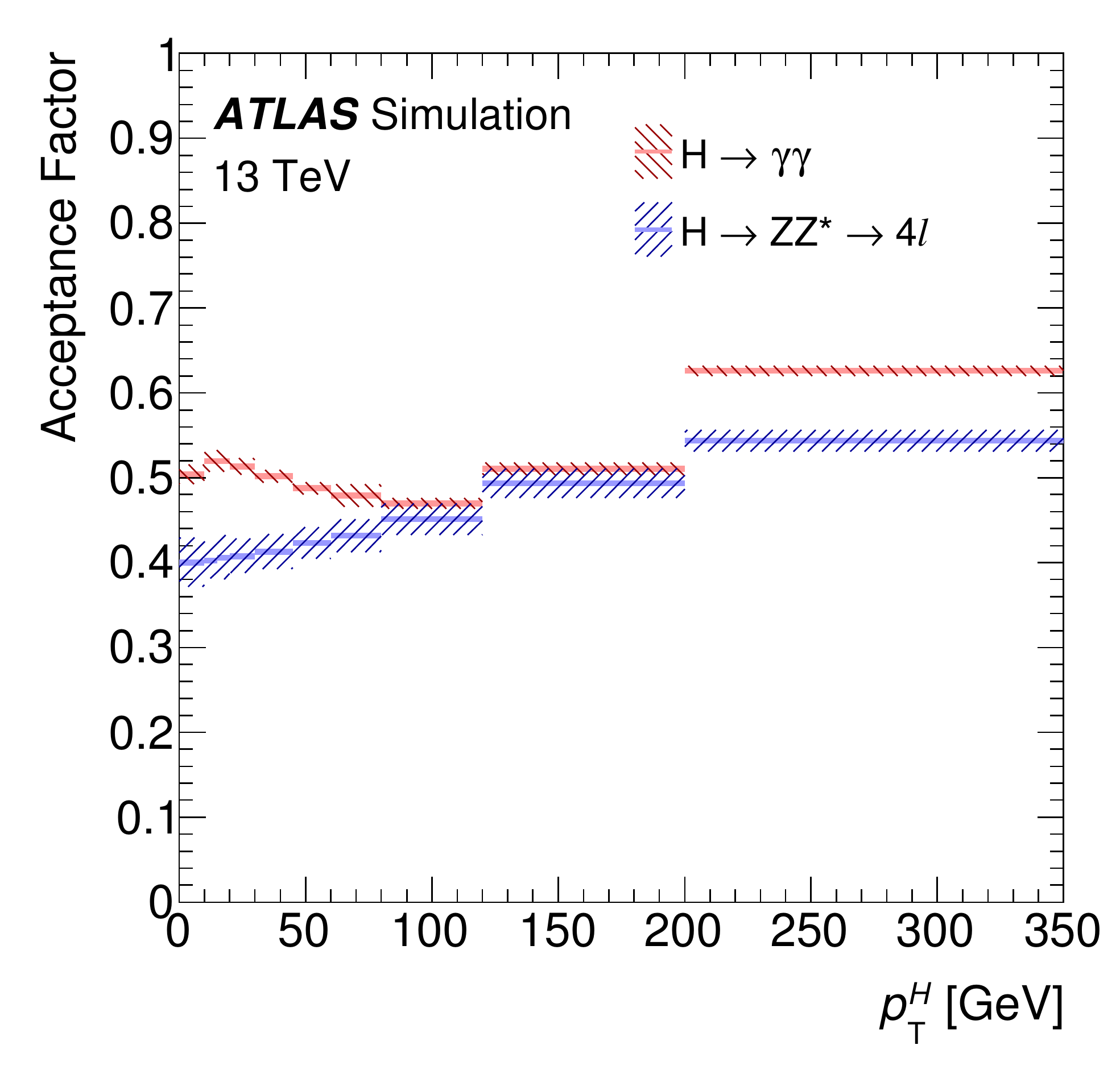}}
  \subfigure[\label{fig:acc_y}]{\includegraphics[width=0.45\linewidth]{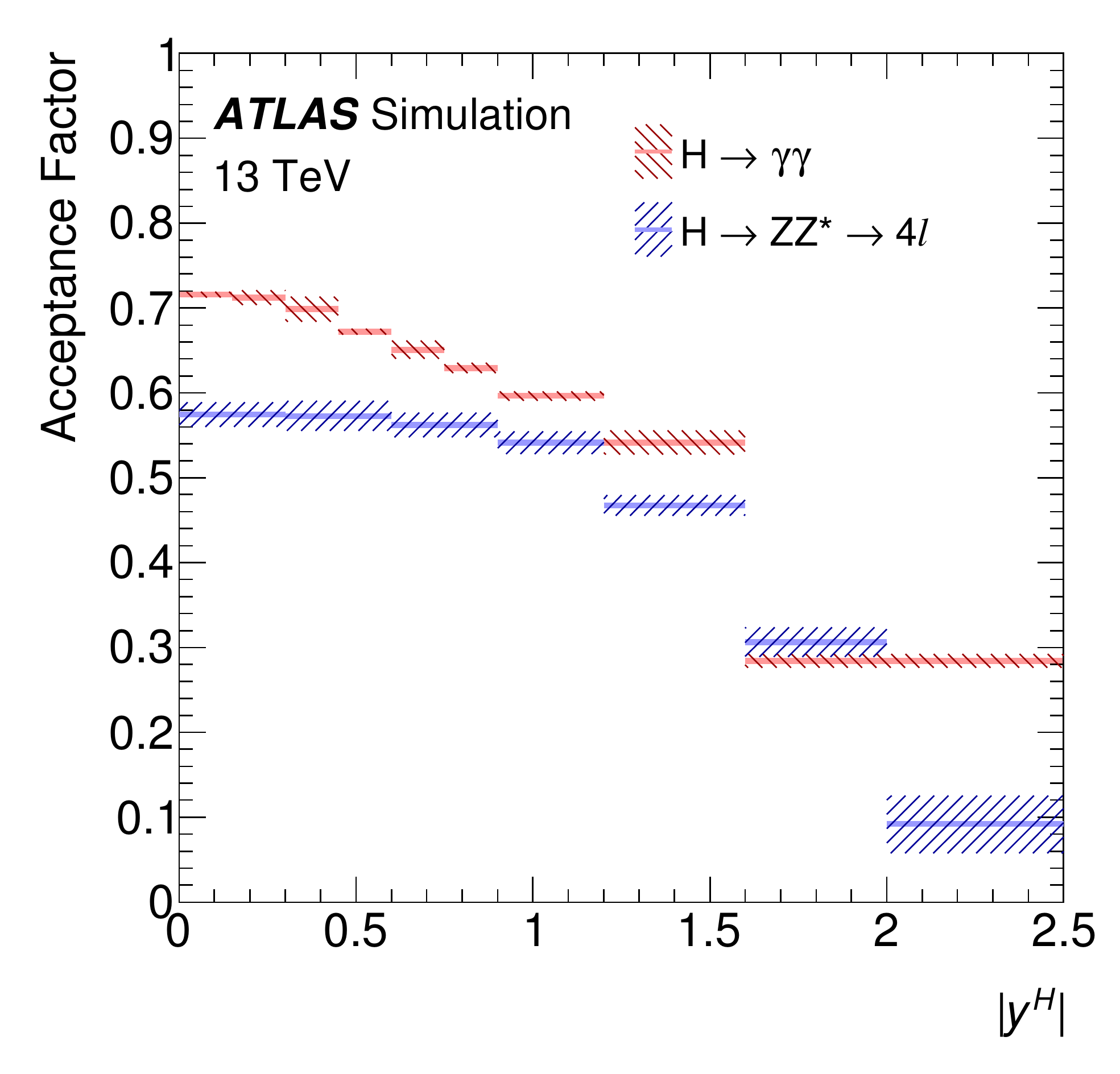}}
  \subfigure[\label{fig:acc_jet1pt}]{\includegraphics[width=0.45\linewidth]{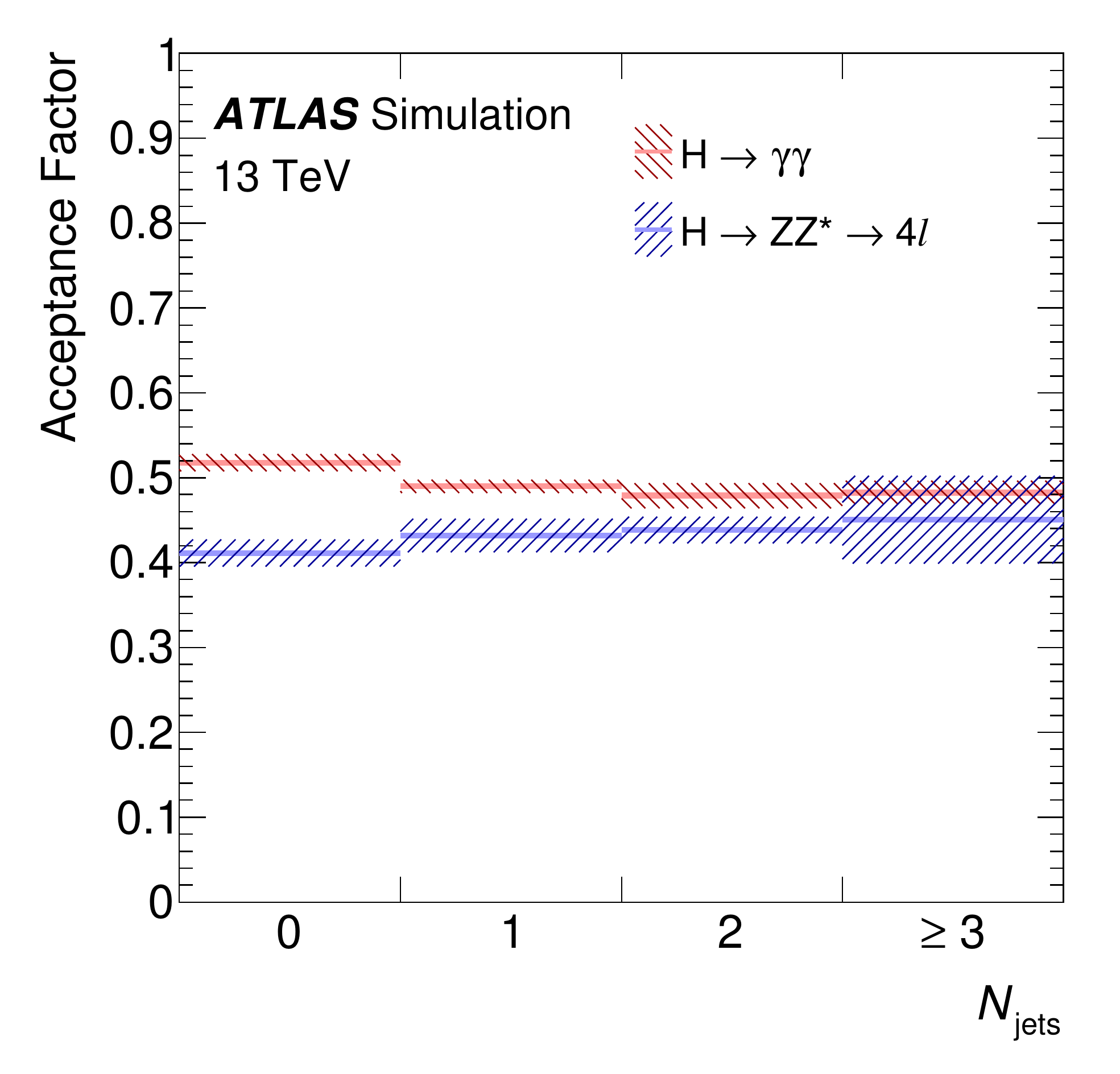}}
  \subfigure[\label{fig:acc_njets}]{\includegraphics[width=0.45\linewidth]{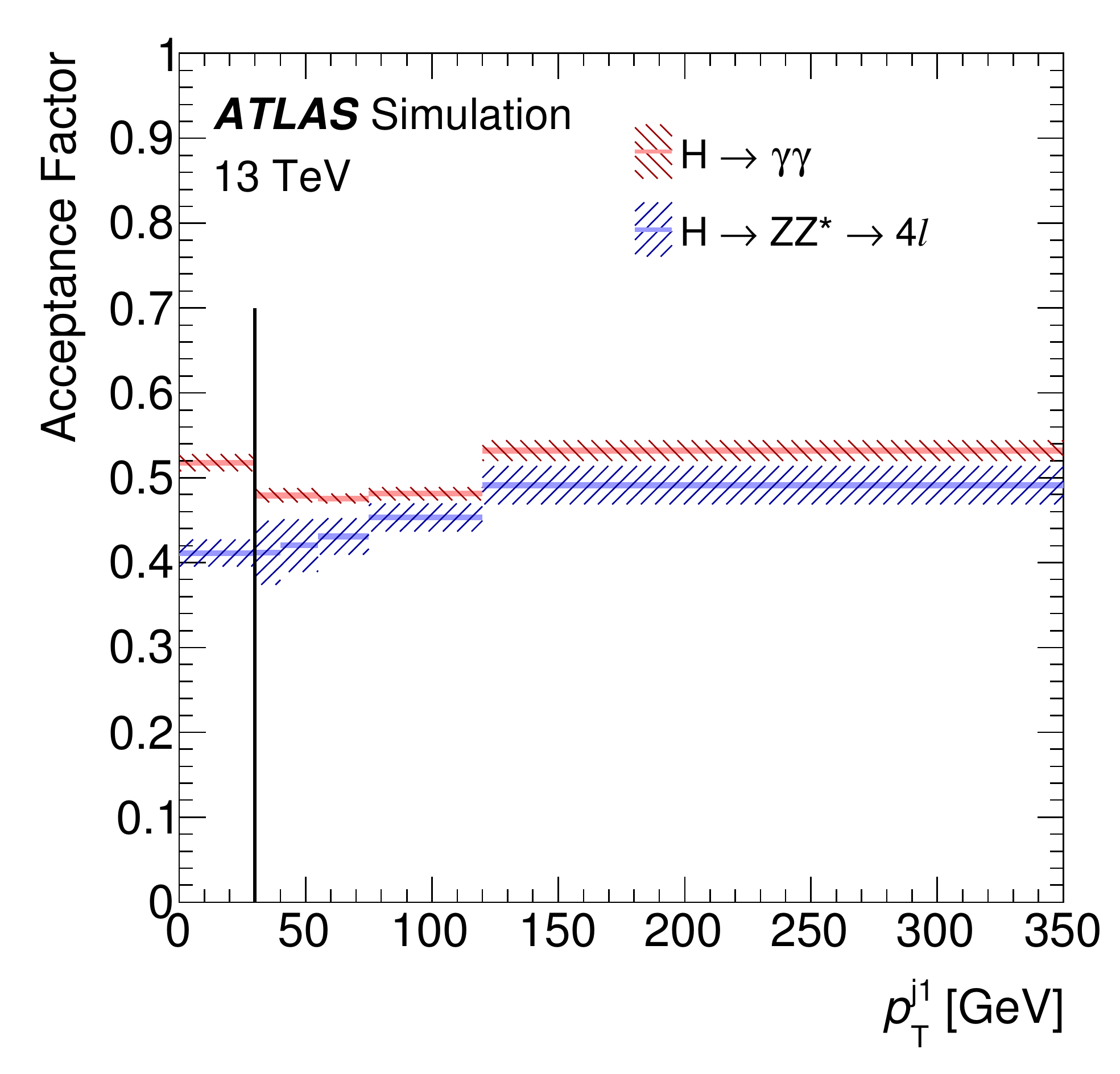}}
 \caption{Acceptance factors for the extrapolation from the fiducial to the total phase space for the \hyy\ decay channel (red) and the \hfourl\ decay channel (blue), for (a) Higgs boson transverse momentum \ptH, (b) Higgs boson rapidity \yH, (c) number of jets \njets\ with \pt\ $>$ 30~\gev, and (d) transverse momentum of the leading jet \jetpt, including systematic uncertainties. The first bin in the \jetpt\ distribution corresponds to the 0-jet bin in the \njets\ distribution, as indicated by the black vertical line. \label{fig:acc}}
\end{figure}

\section{Statistical procedure}
\label{sec:stats}

The combined measurement is based on maximizing the profile-likelihood ratio~\cite{Cowan:2010js}:
\begin{align*}
  \Lambda(\bm{\sigma})=\frac{\mathcal{L}(\bm{\sigma}, \hat{\hat{\bm{\theta}}}(\bm{\sigma}))}{\mathcal{L}(\hat{\bm{\sigma}}, \hat{\bm{\theta}})}.
  \label{eq:likelihood}
\end{align*}

Here $\bm{\sigma}$ are the parameters of interest, $\bm{\theta}$ are the nuisance parameters,
and $\mathcal{L}$ represents the likelihood function.
The $\hat{\bm{\sigma}}$ and $\hat{\bm{\theta}}$ terms denote the unconditional maximum-likelihood estimate of the parameters,
while $\hat{\hat{\bm{\theta}}}(\bm{\sigma})$ is the conditional maximum-likelihood estimate for given parameter values.

The likelihood function $\mathcal{L}$ includes the signal extraction, the correction to particle level, and the extrapolation to the total phase space in each channel.
Therefore, the total cross section as well as the cross sections in different bins for each observable can be derived directly as parameters of interest $\bm{\sigma}$ based on the combined data set from the \hgg\ and \hfourl\ channels.

The distribution shape and normalization systematic uncertainties of all components are included in the likelihood function as nuisance parameters $\bm{\theta}$ with constraints from subsidiary measurements.
This allows the uncertainties to be correlated between bins, decay channels, and correction and acceptance factors.
The uncertainty components of the predicted branching ratios are correlated between the decay channels, as well as the uncertainties in the acceptance and correction factors due to production mode variations, PDF and higher-order corrections, and the parton shower. The uncertainty in the Higgs boson mass, including its effect on the predicted branching ratio, is also correlated between channels.
Experimental uncertainties in the correction factors and the signal extraction in the \hfourl\ decay channel, like the energy scale and resolution of electrons, photons, and jets, and in the luminosity measurement and pileup modelling are also correlated.

The bin boundaries of all probed observables are consistent between the \hgg\ and the \hfourl\ analyses~\cite{HIGG-2016-21,HIGG-2016-25}. Where one bin in one of the measurements corresponds to two bins in the other, the wider bin size is used.
The sum of the cross sections in the finer bins is considered as the parameter of interest in these cases,
and an additional unconstrained nuisance parameter that floats in the fit describes the difference between the merged bins. The normalization and shape uncertainties of the \hgg\ background estimate~\cite{HIGG-2016-21} are fit to the data as nuisance parameters without any initial constraint.

The test statistic $-2 \ln{ \Lambda}$ is assumed to follow a $\chi^2$ distribution for constructing confidence intervals~\cite{Cowan:2010js}. This asymptotic assumption was tested with pseudo-experiments for bins with low numbers of events and found to be appropriate.

The level of agreement between the two channels in the total phase space is evaluated  by using a profiled likelihood as a function of the difference of the cross sections in each bin $i$, $\sigma^i_{\gamma \gamma} - \sigma^i_{4\ell}$. The number of degrees of freedom is the same as the number of bins in the tested distribution. 
The probability that a measured differential cross section is compatible with a theoretical prediction is found by computing a $p$-value based on the difference between the value of $-2 \ln{\Lambda}$ at the unconditional maximum-likelihood estimate and the value obtained by fixing the cross sections in all bins to the ones predicted by the theory.
The uncertainties in the theoretical predictions are ignored when calculating the $p$-values. Including these uncertainties would increase the $p$-values.

\section{Results}
\label{sec:results}

\begin{figure}[tp!]
  \centering
  \subfigure[\label{fig:pt}]{\includegraphics[width=0.45\linewidth]{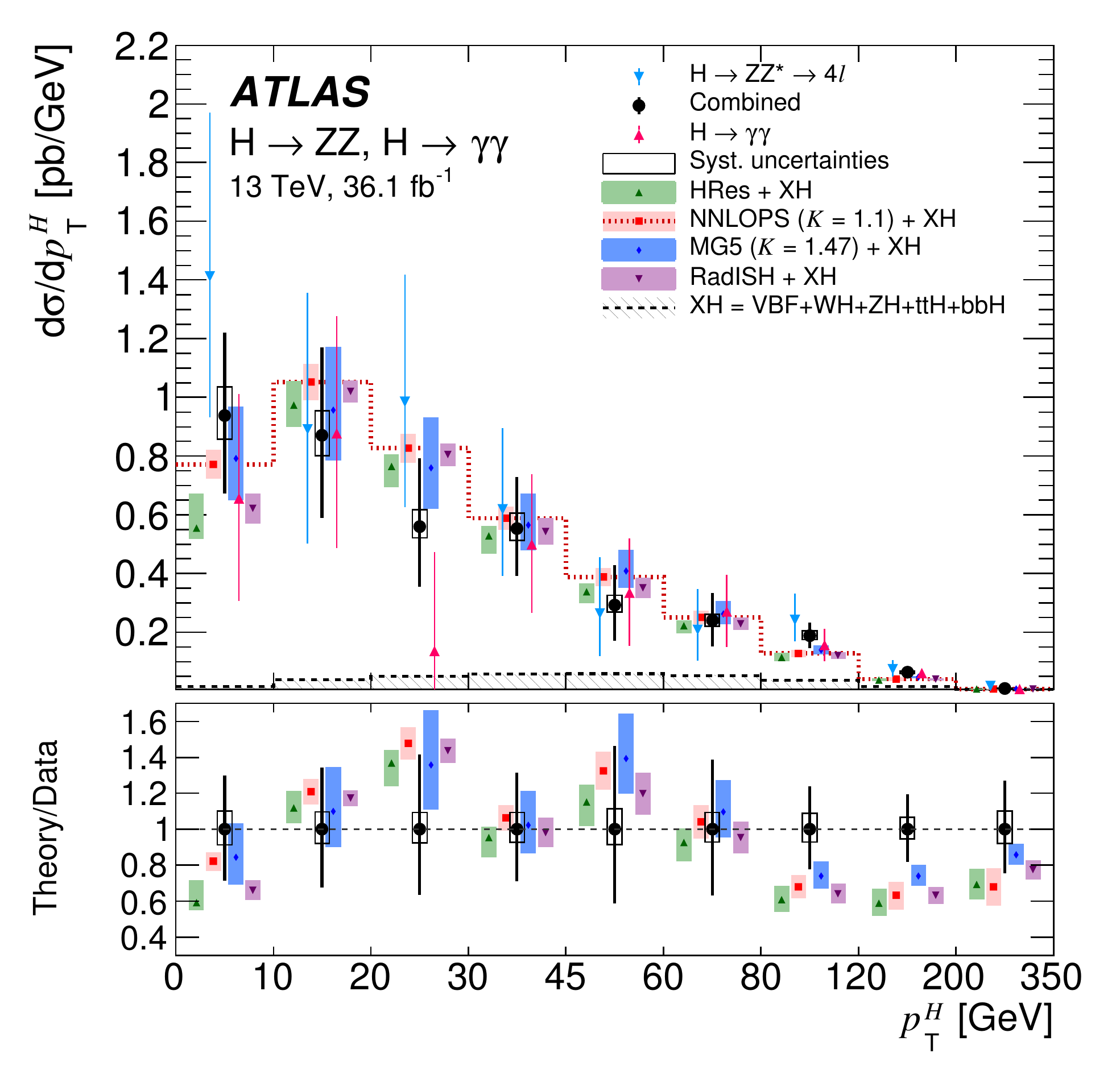}}
  \subfigure[\label{fig:rap}]{\includegraphics[width=0.45\linewidth]{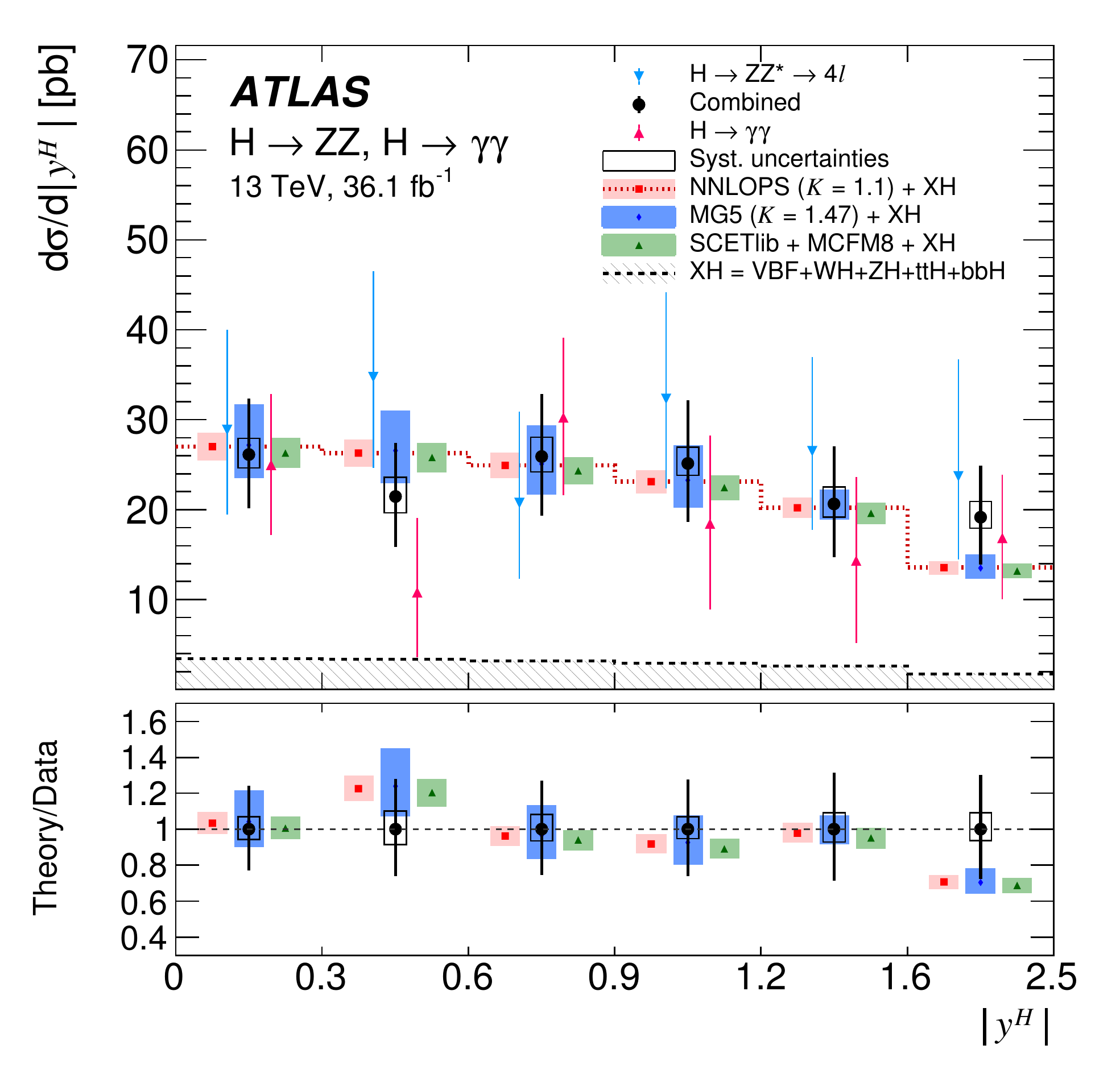}}
  \subfigure[\label{fig:njets}]{\includegraphics[width=0.45\linewidth]{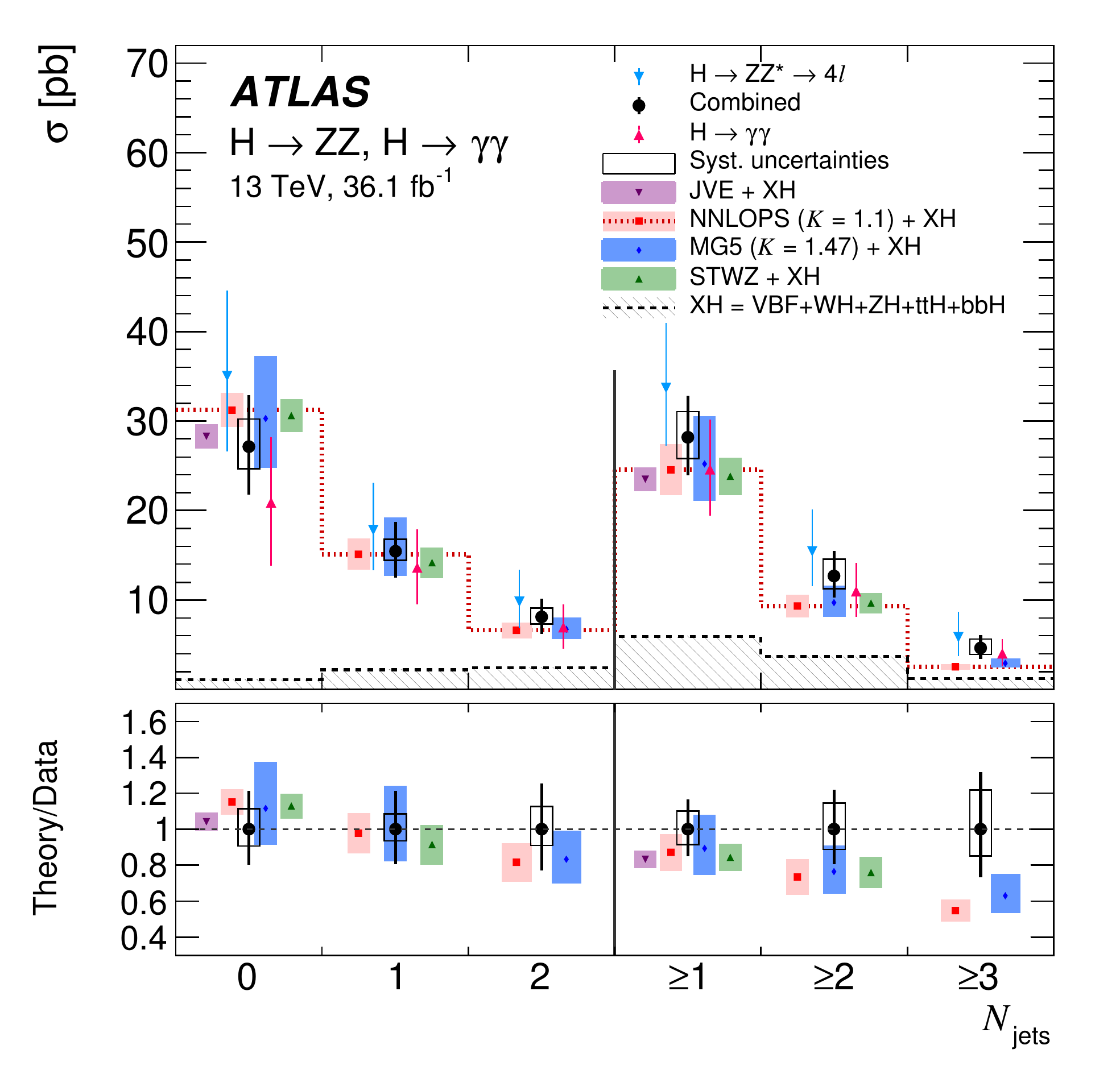}}
  \subfigure[\label{fig:leadingjetpt}]{\includegraphics[width=0.45\linewidth]{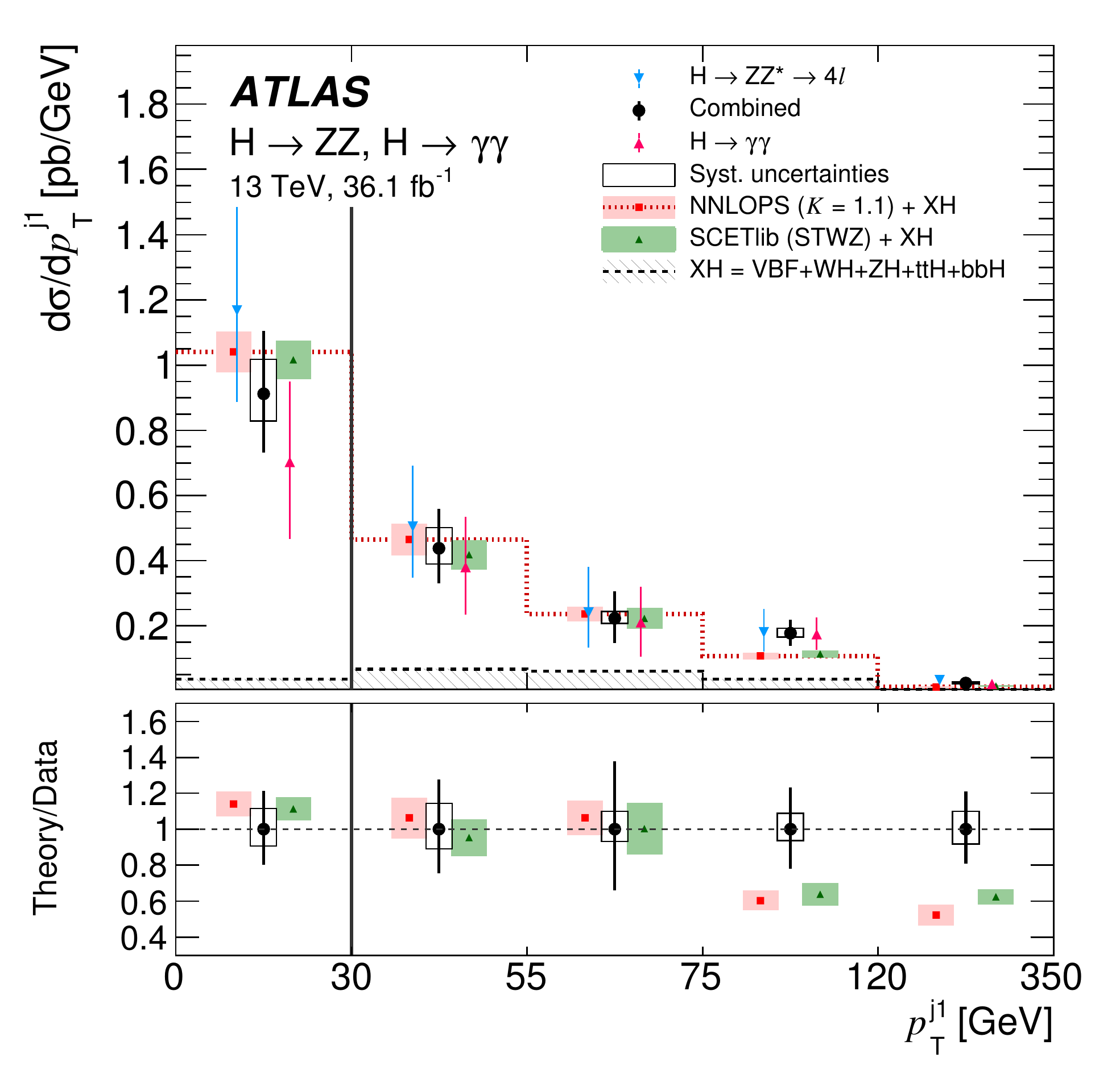}}
  \caption{Differential cross sections in the full phase space measured with the \hgg\ (red upward triangle) and \hfourl\ (blue downward triangle) decay channels, as well as the combined measurement (black circle)  for (a) Higgs boson transverse momentum \ptH, (b) Higgs boson rapidity \yH, (c) number of jets \njets\ with \pt\ $>$ 30~\gev, and (d) the transverse momentum of the leading jet \jetpt. The first bin in the \jetpt\ distribution corresponds to the 0-jet bin in the \njets\ distribution, as indicated by the black vertical line. Different SM predictions are overlaid, their bands indicating the PDF uncertainties as well as uncertainties due to missing higher-order corrections. The ordering of the predictions is the same in the legend as in the figure. Predictions for the other production processes XH are added to the ggF predictions, and also shown separately as a shaded area. The dotted red line corresponds to the central value of the NNLOPS ggF prediction, scaled to the total N$^3$LO cross section by the given $K$-factor, and added to the XH prediction.  The uncertainties due to higher orders in the NNLOPS prediction are obtained as in  Refs.~\cite{HIGG-2016-21,HIGG-2016-25,HIGG-2016-22}.
    The \textsc{Madgraph5\_aMC@NLO} prediction is scaled to the total N$^3$LO cross section by the given $K$-factor.  For better visibility, all bins are shown as having the same size, independent of their numerical width. The panel on the bottom shows the ratio of the predictions to the combined measurement. The total uncertainties of the combined measurement are indicated by the black error bars, the systematic uncertainties by the black open boxes.
    \label{fig:results}}
\end{figure}

The total cross section is measured to be 47.9$^{+9.1}_{-8.6}$~pb in the \hgg\ decay channel and 68$^{+11}_{-10}$~pb in the \hfourl\ channel. The result of the combined measurement is \hbox{57.0 $^{+7.2}_{-6.8}$ ($^{+6.0}_{-5.9}$ (stat.) $^{+4.0}_{-3.3}$ (syst.))~pb}, in agreement with the SM prediction of 55.6$\pm$2.5~pb ~\cite{deFlorian:2016spz}. The results from the individual decay channels are compatible, with a $p$-value of 14\%.

Figure~\ref{fig:results} shows the differential cross sections in the total phase space measured in the \hgg\ and \hfourl\ decay channels as well as the combined measurement as a function of \ptH, \yH, \njets, and \jetpt. Different SM predictions are overlaid. The uncertainties in the \textsc{Madgraph5\_aMC@NLO} distribution are larger than for the other predictions, as this prediction is at NLO accuracy only.

For all differential observables and bins, the measurement is dominated by statistical uncertainties, which vary between 20\% and 30\%.
Significant uncertainties affecting all observables, including the total cross section, include the uncertainty in the 2015 and 2016 integrated luminosity, which is 3.2\%~\cite{Aaboud:2016hhf}, affecting the signal and simulated background estimates in the \hfourl\ decay channel, with an impact of about 4\% on the measurement, and the background estimate in the \hgg\ signal extraction~\cite{HIGG-2016-21}, typically 2$-$6\%. For \njets\ and \jetpt, the uncertainties in the reconstruction of the jet energy scale and resolution are important as well, typically 3$-$6\% ($>$10\% for \njets $\geq$ 3)~\cite{Aaboud:2017jcu}.

The level of agreement between the two channels in the total phase space is quantified by the corresponding $p$-values: 58\% for \ptH, 40\% for \yH, 53\% for \njets\ and 67\% for \jetpt.

Table~\ref{tab:comptheory} shows the $p$-values indicating reasonable agreement between the probed SM predictions and the measurement. The relatively low $p$-value for \progname{HRes} can be explained by the lower computed total cross section, as this prediction is at NNLO+NNLL accuracy. 
The lower $p$-values for \jetpt\ reflect the lower predictions compared to the measurement for high jet \pt. Compatibility checks of individual bins indicate less than 3$\sigma$ local discrepancy.

\begin{table}
\centering
\caption{$p$-values in percent indicating the probabilities that the measured differential cross sections are compatible with various SM ggF predictions. The NNLOPS and \textsc{Madgraph5\_aMC@NLO} predictions are scaled to the total N$^3$LO cross section by the given $K$-factors. The non-ggF predictions are added, as discussed in Section~\ref{sec:montecarlo}. The uncertainties in the theoretical predictions are ignored when calculating the $p$-values.
\label{tab:comptheory} }
\vspace{0.2cm}
\begin{tabular}{lcccc}
  \hline\hline
   		$p$-values [\%]		& \ptH   & \yH    & \njets & \jetpt \\
  \hline
  NNLOPS ($K = 1.1$)		         & 29 	& 92 	&43	 & ~6 \\
  \textsc{HRes}                          & 16 & -- & -- & -- \\
  \textsc{RaDISH + NNLOJET}              & 30  & -- & -- & -- \\
  \textsc{SCETlib}                       & -- & 91 & -- & 23 \\
  \textsc{Madgraph5\_aMC@NLO} ($K = 1.47$) & 77 & 91 & 65 & -- \\
  \hline
  \hline
\end{tabular}
\end{table}

\FloatBarrier

\section{Conclusion}

A combined measurement of the total and differential cross sections in the \hgg\ and \hfourl\ decay channels was performed, using 36.1~fb$^{-1}$ of 13~\tev\ proton--proton collision data produced by the LHC and recorded by the ATLAS detector in 2015 and 2016.
Good agreement is observed when comparing the results from the two channels, extrapolated to a common phase space.
The total Higgs boson production cross section is measured to be \hbox{57.0$^{+6.0}_{-5.9}$ (stat.) $^{+4.0}_{-3.3}$ (syst.)~pb}, in agreement with the Standard Model prediction. 
Differential cross-section measurements are presented for the Higgs boson transverse momentum distribution, Higgs boson rapidity, number of jets produced together with the Higgs boson, and the transverse momentum of the leading jet.
The larger data set and the combination of the two decay channels give measurement uncertainties that are significantly smaller than in previous results.
The combined results agree with Standard Model predictions.

\section*{Acknowledgements}


We thank CERN for the very successful operation of the LHC, as well as the
support staff from our institutions without whom ATLAS could not be
operated efficiently.

We acknowledge the support of ANPCyT, Argentina; YerPhI, Armenia; ARC, Australia; BMWFW and FWF, Austria; ANAS, Azerbaijan; SSTC, Belarus; CNPq and FAPESP, Brazil; NSERC, NRC and CFI, Canada; CERN; CONICYT, Chile; CAS, MOST and NSFC, China; COLCIENCIAS, Colombia; MSMT CR, MPO CR and VSC CR, Czech Republic; DNRF and DNSRC, Denmark; IN2P3-CNRS, CEA-DRF/IRFU, France; SRNSFG, Georgia; BMBF, HGF, and MPG, Germany; GSRT, Greece; RGC, Hong Kong SAR, China; ISF, I-CORE and Benoziyo Center, Israel; INFN, Italy; MEXT and JSPS, Japan; CNRST, Morocco; NWO, Netherlands; RCN, Norway; MNiSW and NCN, Poland; FCT, Portugal; MNE/IFA, Romania; MES of Russia and NRC KI, Russian Federation; JINR; MESTD, Serbia; MSSR, Slovakia; ARRS and MIZ\v{S}, Slovenia; DST/NRF, South Africa; MINECO, Spain; SRC and Wallenberg Foundation, Sweden; SERI, SNSF and Cantons of Bern and Geneva, Switzerland; MOST, Taiwan; TAEK, Turkey; STFC, United Kingdom; DOE and NSF, United States of America. In addition, individual groups and members have received support from BCKDF, the Canada Council, CANARIE, CRC, Compute Canada, FQRNT, and the Ontario Innovation Trust, Canada; EPLANET, ERC, ERDF, FP7, Horizon 2020 and Marie Sk{\l}odowska-Curie Actions, European Union; Investissements d'Avenir Labex and Idex, ANR, R{\'e}gion Auvergne and Fondation Partager le Savoir, France; DFG and AvH Foundation, Germany; Herakleitos, Thales and Aristeia programmes co-financed by EU-ESF and the Greek NSRF; BSF, GIF and Minerva, Israel; BRF, Norway; CERCA Programme Generalitat de Catalunya, Generalitat Valenciana, Spain; the Royal Society and Leverhulme Trust, United Kingdom.

The crucial computing support from all WLCG partners is acknowledged gratefully, in particular from CERN, the ATLAS Tier-1 facilities at TRIUMF (Canada), NDGF (Denmark, Norway, Sweden), CC-IN2P3 (France), KIT/GridKA (Germany), INFN-CNAF (Italy), NL-T1 (Netherlands), PIC (Spain), ASGC (Taiwan), RAL (UK) and BNL (USA), the Tier-2 facilities worldwide and large non-WLCG resource providers. Major contributors of computing resources are listed in Ref.~\cite{ATL-GEN-PUB-2016-002}.

\printbibliography

\clearpage
 
\begin{flushleft}
{\Large The ATLAS Collaboration}

\bigskip

M.~Aaboud$^\textrm{\scriptsize 34d}$,    
G.~Aad$^\textrm{\scriptsize 99}$,    
B.~Abbott$^\textrm{\scriptsize 124}$,    
O.~Abdinov$^\textrm{\scriptsize 13,*}$,    
B.~Abeloos$^\textrm{\scriptsize 128}$,    
D.K.~Abhayasinghe$^\textrm{\scriptsize 91}$,    
S.H.~Abidi$^\textrm{\scriptsize 164}$,    
O.S.~AbouZeid$^\textrm{\scriptsize 143}$,    
N.L.~Abraham$^\textrm{\scriptsize 153}$,    
H.~Abramowicz$^\textrm{\scriptsize 158}$
% ...
\bigskip
\\

% ...
$^{13}$Institute of Physics, Azerbaijan Academy of Sciences, Baku; Azerbaijan.\\
% ...
$^{34}$$^{(a)}$Facult\'e des Sciences Ain Chock, R\'eseau Universitaire de Physique des Hautes Energies - Universit\'e Hassan II, Casablanca;$^{(b)}$Centre National de l'Energie des Sciences Techniques Nucleaires (CNESTEN), Rabat;$^{(c)}$Facult\'e des Sciences Semlalia, Universit\'e Cadi Ayyad, LPHEA-Marrakech;$^{(d)}$Facult\'e des Sciences, Universit\'e Mohamed Premier and LPTPM, Oujda;$^{(e)}$Facult\'e des sciences, Universit\'e Mohammed V, Rabat; Morocco.\\
% ...
$^{91}$Department of Physics, Royal Holloway University of London, Egham; United Kingdom.\\
% ...
$^{99}$CPPM, Aix-Marseille Universit\'e, CNRS/IN2P3, Marseille; France.\\
% ...
$^{124}$Homer L. Dodge Department of Physics and Astronomy, University of Oklahoma, Norman OK; United States of America.\\
% ...
$^{128}$LAL, Universit\'e Paris-Sud, CNRS/IN2P3, Universit\'e Paris-Saclay, Orsay; France.\\
% ...
$^{143}$Santa Cruz Institute for Particle Physics, University of California Santa Cruz, Santa Cruz CA; United States of America.\\
% ...
$^{153}$Department of Physics and Astronomy, University of Sussex, Brighton; United Kingdom.\\
% ...
$^{158}$Raymond and Beverly Sackler School of Physics and Astronomy, Tel Aviv University, Tel Aviv; Israel.\\
% ...
$^{164}$Department of Physics, University of Toronto, Toronto ON; Canada.\\
% ...
$^{a}$ Also at  Department of Physics, University of Malaya, Kuala Lumpur; Malaysia.\\
$^{b}$ Also at Borough of Manhattan Community College, City University of New York, NY; United States of America.\\
$^{c}$ Also at California State University, East Bay; United States of America.\\
$^{d}$ Also at Centre for High Performance Computing, CSIR Campus, Rosebank, Cape Town; South Africa.\\
$^{e}$ Also at CERN, Geneva; Switzerland.\\
% ...
$^{*}$ Deceased

\end{flushleft}


--------------------------------------------

\end{document}